\documentclass[12pt]{article}
\usepackage{times}
\usepackage{geometry}
\usepackage[utf8]{inputenc}
\usepackage{enumitem,amssymb}
\usepackage{ragged2e}
\newlist{thematic}{itemize}{8}
\setlist[thematic]{label=$\square$}
\usepackage{pifont}

\usepackage{graphics,graphicx}
\usepackage{sidecap}
 \usepackage{wrapfig, framed, caption}
    \usepackage{lipsum}%
\usepackage{bm}
\def\chandra{{\it Chandra\/}}
\def\xmm{{XMM-{\it Newton\/}}}

\usepackage{sidecap}

\begin{document}
\raggedright
\huge
Astro2020 Science White Paper \linebreak

Cosmic evolution of supermassive black holes: A view into the next two decades \linebreak

\normalsize

\noindent \textbf{Thematic Areas:} \hspace*{60pt} $\square$ Planetary Systems \hspace*{10pt} $\square$ Star and Planet Formation \hspace*{20pt}\linebreak
$\boxtimes$ Formation and Evolution of Compact Objects \hspace*{31pt} $\boxtimes$ Cosmology and Fundamental Physics \linebreak
  $\square$ Stars and Stellar Evolution \hspace*{1pt} $\square$ Resolved Stellar Populations and their Environments \hspace*{40pt} \linebreak
  $\boxtimes$ Galaxy Evolution   \hspace*{45pt} $\square$ Multi-Messenger Astronomy and Astrophysics \hspace*{65pt} \linebreak
 
\justify
\newcommand{\Amherst}{University of Massachusetts, Amherst, MA 01003 USA}
\newcommand{\ANLHEP}{HEP Division, Argonne National Laboratory, Lemont, IL 60439, USA}
\newcommand{\APC}{Laboratoire Astroparticule et Cosmologie (APC), CNRS/IN2P3, Universit\'e Paris Diderot, 10, rue Alice Domon et Léonie Duquet, 75205 Paris Cedex 13, France}
\newcommand{\ASU}{Arizona State University, Tempe, AZ  85287, USA}
\newcommand{\BenGurion}{Department of Physics, Ben-Gurion University, Be'er Sheva 84105, Israel}
\newcommand{\BNL}{Brookhaven National Laboratory, Upton, NY 11973, USA}
\newcommand{\Brown}{Brown University, Providence, RI 02912, USA}
\newcommand{\Bub}{Boston University, Boston, MA 02215, USA}
\newcommand{\BU}{Boston University, Boston, MA 02215, USA}
\newcommand{\Buffalo}{Department of Physics, University at Buffalo, SUNY Buffalo, NY 14260 USA}
\newcommand{\Caltech}{California Institute of Technology, Pasadena, CA 91125, USA}
\newcommand{\Cardiff}{School of Physics and Astronomy, Cardiff University, The Parade, Cardiff, CF24 3AA, UK}
\newcommand{\Carleton}{Carleton University, K1S 5B6 Ottawa, Canada}
\newcommand{\Carnegie}{The Observatories of the Carnegie Institution for Science, 813 Santa Barbara St., Pasadena, CA 91101, USA}
\newcommand{\Cavendish}{Astrophysics Group, Cavendish Laboratory, J.J.Thomson Avenue, Cambridge, CB3 0HE, UK}
\newcommand{\CCA}{Center for Computational Astrophysics, 162 5th Ave, 10010, New York, NY, USA}
\newcommand{\CPPM}{Aix Marseille Univ, CNRS/IN2P3, CPPM, Marseille, France}
\newcommand{\CEADAP}{D\'epartement d’Astrophysique, CEA Saclay DSM/Irfu, 91191 Gif-sur-Yvette, France}
\newcommand{\CERN}{CERN, Geneva, Switzerland}
\newcommand{\CfA}{Center for Astrophysics | Harvard \& Smithsonian, 60 Garden st, Cambridge, MA 02138, USA}
\newcommand{\CFT}{Center for Theoretical Physics, Polish Academy of Sciences, al. Lotnik\'{o}w 32/46, 02-668, Warsaw, Poland}
\newcommand{\Cincinnati}{University of Cincinnati, Cincinnati, OH 45221, USA}
\newcommand{\CITA}{Canadian Institute for Theoretical Astrophysics, University of Toronto, Toronto, ON M5S 3H8, Canada}
\newcommand{\CNRSA}{CNRS, Laboratoire d'Annecy-le-Vieux de Physique Th\'{e}orique, France}
\newcommand{\CNYang}{C.N. Yang Institute for Theoretical Physics State University of New York Stony Brook, NY 11794, USA}
\newcommand{\CMUCosmo}{Department 
of Physics, McWilliams Center for Cosmology, Carnegie Mellon University, USA}
\newcommand{\Columbia}{Columbia University, New York, NY 10027, USA}
\newcommand{\Cornell}{Cornell University, Ithaca, NY 14853, USA}
\newcommand{\CPthree}{CP3-Origins, 5230 Odense, Denmark}
\newcommand{\CWRU}{Case Western Reserve University, Cleveland, OH 44106, USA}
\newcommand{\daa}{Department of Astronomy and Astrophysics, University of Toronto, ON, M5S3H4, Canada}
\newcommand{\damtp}{DAMTP, Centre for Mathematical Sciences, Wilberforce Road, Cambridge, UK, CB3 0WA}
\newcommand{\DESY}{DESY,  22607 Hamburg, Germany}
\newcommand{\DFI}{Departamento de F\'isica, FCFM, Universidad de Chile, Blanco Encalada 2008, Santiago, Chile}
\newcommand{\DOE}{US. Department of Energy, Germantown, MD 20874, USA}
\newcommand{\drexel}{Drexel University, Philadelphia, PA 19104, USA}
\newcommand{\Duke}{Duke University and Triangle Universitites Nuclear Laboratory, Durham, NC 27708, USA}
\newcommand{\DukePhys}{Department of Physics, Duke University, Durham, NC 27708, USA}
\newcommand{\dunlap}{Dunlap Institute for Astronomy and Astrophysics, University of Toronto, ON, M5S3H4, USA}
\newcommand{\Durham}{Department of Physics, Lower Mountjoy, South Rd, Durham DH1 3LE, UK}
\newcommand{\ED}{University of Edinburgh, EH8 9YL Edinburgh, UK}
\newcommand{\EPFL}{Institute of Physics, Laboratory of Astrophysics, Ecole Polytechnique Fédérale de Lausanne (EPFL), Observatoire de Sauverny, 1290 Versoix, Switzerland}
\newcommand{\ETH}{ETH Zurich, Institute for Particle Physics, 8093 Zurich, Switzerland}
\newcommand{\FNAL}{Fermi National Accelerator Laboratory, Batavia, IL 60510, USA}
\newcommand{\FQAUB}{Dept. de F\' isica Qu\` antica i Astrof\' isica, Universitat de Barcelona, Mart\' i i Franqu\` es 1, E08028 Barcelona, Spain}
\newcommand{\FSU}{Florida State University, Tallahassee, FL 32306, USA}
\newcommand{\Glasgow}{University of Glasgow, G12 8QQ Glasgow, UK}
\newcommand{\GRAPPA}{GRAPPA Institute, University of Amsterdam, Science Park 904, 1098 XH Amsterdam, The Netherlands}
\newcommand{\GSFC}{Goddard Space Flight Center, Greenbelt, MD 20771 USA}
\newcommand{\GWU}{George Washington University, Washington, DC 20052, USA}
\newcommand{\Hampton}{Hampton University, Hampton, VA 23668, USA}
\newcommand{\HarvardPhys}{Department of Physics, Harvard University, Cambridge, MA 02138, USA}
\newcommand{\Haverford}{Haverford College, 370 Lancaster Ave, Haverford PA, 19041, USA}
\newcommand{\Hawaii}{University of Hawaii, Honolulu, HI 96822, USA}
\newcommand{\HKUST}{The Hong Kong University of Science and Technology, Hong Kong SAR, China}
\newcommand{\houston}{University of Houston, Houston, TX 77204, USA}
\newcommand{\IAP}{Institut d'Astrophysique de Paris (IAP), CNRS \& Sorbonne University, Paris, France}
\newcommand{\IAS}{Institute for Advanced Study, Princeton, NJ 08540, USA}
\newcommand{\IBS}{Institute for Basic Science (IBS), Daejeon 34051, Korea}
\newcommand{\ICC}{ICC, University of Barcelona, IEEC-UB, Mart\' i i Franqu\` es, 1, E08028 Barcelona, Spain}
\newcommand{\ICCD}{Institute for Computational Cosmology, Department of Physics, Durham University, South Road, Durham, DH1 3LE, UK}
\newcommand{\ICE}{Institute of Space Sciences, Campus UAB, Carrer de Can Magrans, s/n, 08193 Barcelona, Spain}
\newcommand{\ICRR}{Institute for Cosmic Ray Resaerch, The University of Tokyo, 456 Higashi-Mozumi, Kamioka, Hida, Gifu 506-1205, Japan}
\newcommand{\ICTP}{International Centre for Theoretical Physics, Strada Costiera, 11, I-34151 Trieste, Italy}
\newcommand{\IFAE}{Institut de Fisica d’Altes Energies, The Barcelona Institute of Science and Technology, Campus UAB, 08193 Bellaterra (Barcelona), Spain}
\newcommand{\IFPU}{IFPU - Institute for Fundamental Physics of the Universe, Via Beirut 2, 34014 Trieste, Italy}
\newcommand{\IFT}{Instituto de Fisica Teorica UAM/CSIC, Universidad Autonoma de Madrid, 28049 Madrid, Spain}
\newcommand{\IFUNAM}{IFUNAM - Instituto de F\'{i}sica, Universidad Nacional Aut\'onoma de M\'etico, 04510 CDMX, M\'exico}
\newcommand{\IHEP}{Institute of High Energy Physics, Austrian Academy of Sciences, 1050 Vienna, Austria}
\newcommand{\Imperial}{Theoretical Physics, Blackett Laboratory, Imperial College, London, SW7 2AZ, UK}
\newcommand{\Indiana}{Indiana University, Bloomington, IN 47405, USA}
\newcommand{\INAFOATs}{INAF - Osservatorio Astronomico di Trieste, Via G.B. Tiepolo 11, 34143 Trieste, Italy}
\newcommand{\INAFOAS}{INAF - Osservatorio di Astrofisica e Scienza dello Spazio, via Piero Gobetti 93/3, I-40129 Bologna, Italy}
\newcommand{\INFNCag}{Istituto Nazionale di Fisica Nucleare, Sezione di Cagliari,  09126 Cagliari, Italy}
\newcommand{\INFNCat}{Istituto Nazionale di Fisica Nucleare, Sezione di Catania, 95125 Catania, Italy}
\newcommand{\INFNG}{Istituto Nazionale di Fisica Nucleare, Sezione di Genova, 16146 Genova, Italy}
\newcommand{\INFN}{INFN – National Institute for Nuclear Physics, Via Valerio 2, I-34127 Trieste, Italy}
\newcommand{\INFNFE}{Istituto Nazionale di Fisica Nucleare, Sezione di Ferrara, 40122, Italy }
\newcommand{\INFNLNF}{Istituto Nazionale di Fisica Nucleare, Laboratori Nazionali di Frascati, 00044 Frascati, Italy}
\newcommand{\INFNLNS}{Istituto Nazionale di Fisica Nucleare, Laboratori Nazionali del Sud, 95125 Catania, Italy}
\newcommand{\INFNN}{Istituto Nazionale di Fisica Nucleare, Sezione di Napoli, 80125 Napoli, Italy }
\newcommand{\INFNRM}{Istituto Nazionale di Fisica Nucleare, Sezione di Roma, 00185 Roma, Italy}
\newcommand{\INFNT}{Istituto Nazionale di Fisica Nucleare, Sezione di Torino, 10125, Italy }
\newcommand{\ioa}{Institute of Astronomy, University of Cambridge,Cambridge CB3 0HA, UK}
\newcommand{\IPP}{Institute for Particle Physics, BC V8W 3P6 Victoria, Canada}
\newcommand{\IPMU}{Kavli Insitute for the Physics and Mathematics of the Universe (WPI), University of Tokyo, 277-8583 Kashiwa , Japan}
\newcommand{\IPNL}{Universit\'e de Lyon, F-69622, Lyon, France; Universit\'e de Lyon 1, Villeurbanne; CNRS/IN2P3, Institut de Physique Nucl\'eaire de Lyon}
\newcommand{\IRFU}{IRFU, CEA, Universit\'e Paris-Saclay, F-91191 Gif-sur-Yvette, France}
\newcommand{\ITFA}{Institute for Theoretical Physics, University of Amsterdam, Science Park 904, 1098 XH Amsterdam, The Netherlands}
\newcommand{\IUCAA}{The Inter-University Centre for Astronomy and Astrophysics, Pune, 411007, India}
\newcommand{\Jerusalem}{Hebrew University of Jerusalem, 91904 Jerusalem, Israel}
\newcommand{\JHU}{Johns Hopkins University, Baltimore, MD 21218, USA}
\newcommand{\JLAB}{Thomas Jefferson National Laboratory, Newport News, VA 23606, USA}
\newcommand{\JPL}{Jet Propulsion Laboratory, California Institute of Technology, Pasadena, CA, USA}
\newcommand{\KASSI}{Korea Astronomy and Space Science Institute, Daejeon 34055, Korea}
\newcommand{\kavli}{Kavli Institute for Cosmology, Cambridge, UK, CB3 0HA}
\newcommand{\KIAS}{School of Physics, Korea Institute for Advanced Study, 85 Hoegiro, Dongdaemun-gu, Seoul 130-722, Korea}
\newcommand{\KICP}{Kavli Institute for Cosmological Physics, Chicago, IL 60637, USA}
\newcommand{\KIPAC}{Kavli Institute for Particle Astrophysics and Cosmology, Stanford 94305, USA}
\newcommand{\KINGS}{King's College London, WC2R 2LS London, UK}
\newcommand{\Kobe}{Kobe University, 657-8501 Kobe, Japan}
\newcommand{\KPH}{Johannes Gutenberg University, 55128 Mainz, Germany}
\newcommand{\KPMU}{University of Tokyo, 277-8583  Kashiwa , Japan}
\newcommand{\KSU}{Kansas State University, Manhattan, KS 66506, USA}
\newcommand{\Lafayette}{Lafayette College, Easton, PA 18042, USA}
\newcommand{\LANL}{Los Alamos National Laboratory, Los Alamos, NM 87545, USA}
\newcommand{\LBL}{Lawrence Berkeley National Laboratory, Berkeley, CA 94720, USA}
\newcommand{\Leiden}{Lorentz Institute, Leiden University, Niels Bohrweg 2,Leiden, NL 2333 CA, The Netherlands}
\newcommand{\Liverpool}{University of Liverpool,  L69 7ZE Liverpool , UK}
\newcommand{\LLNL}{Lawrence Livermore National Laboratory, Livermore, CA, 94550, USA}
\newcommand{\LPC}{Universit\'e Clermont Auvergne, CNRS/IN2P3, Laboratoire de Physique de Clermont, F-63000 Clermont-Ferrand, France}
\newcommand{\LPNHE}{Sorbonne Universit\'e, Universit\'e Paris Diderot, CNRS/IN2P3, Laboratoire de Physique Nucl\'eaire et de Hautes Energies, LPNHE, 4 Place Jussieu, F-75252 Paris, France}
\newcommand{\McGill}{McGill University, Montreal, QC H3A 2T8, Canada}
\newcommand{\Melbourne}{School of Physics, The University of Melbourne, Parkville, VIC 3010, Australia}
\newcommand{\Mines}{Colorado School of Mines, Golden, CO 80401, USA}
\newcommand{\MIT}{Massachusetts Institute of Technology, Cambridge, MA 02139, USA}
\newcommand{\MPE}{Max-Planck-Institut f\"{u}r extraterrestrische Physik (MPE), Giessenbachstrasse 1, D-85748 Garching bei M\"unchen, Germany}
\newcommand{\MPIA}{Max-Planck-Institut f\"{u}r Astronomie, K\"{o}nigstuhl 17, D-69117, Heidelberg, Germany}
\newcommand{\MPP}{Max-Planck-Institut f\"{u}r Physik (Werner-Heisenberg-Institut), F\"ohringer Ring 6, D-80805 M\"unchen, Germany}
\newcommand{\LUPM}{Laboratoire Univers et Particules de Montpellier, Univ. Montpellier and CNRS, 34090 Montpellier, France}
\newcommand{\NAOC}{National Astronomical Observatories, Chinese Academy of Sciences, PR China}
\newcommand{\NCBJ}{National Center for Nuclear Research, Ul.Pasteura 7,Warsaw, Poland}
\newcommand{\NCU}{National Central University, Taoyuan City 32001, Taiwan (R.O.C.)}
\newcommand{\NCSU}{Physics Department, North Carolina State Universitym, 2401 Stinson Dr, Raleigh, NC 27607, USA}
\newcommand{\ND}{University of Notre Dame,vNotre Dame, IN 46556, USA}
\newcommand{\NIU}{Northern Illinois University, DeKalb, Illinois 60115, USA}
\newcommand{\NMSU}{New Mexico State University, Las Cruces, NM 88003, USA}
\newcommand{\NOAO}{National Optical Astronomy Observatory, 950 N. Cherry Ave., Tucson, AZ 85719 USA}
\newcommand{\Northwestern}{Northwestern University, Evanston, IL 60201, USA}
\newcommand{\Nottingham}{University of Nottingham, NG7 2RD Nottingham, UK}
\newcommand{\NWU}{Northwestern University, Evanston, IL 60208, USA}
\newcommand{\NYU}{New York University, New York, NY 10003, USA}
\newcommand{\OK}{ University of Oklahoma, Norman, OK 73019, USA}
\newcommand{\ORNL}{Oak Ridge National Laboratory, Oak Ridge, TN 37831, USA}
\newcommand{\OSU}{The Ohio State University, Columbus, OH 43212, USA}
\newcommand{\OU}{Department of Physics and Astronomy, Ohio University, Clippinger Labs, Athens, OH 45701, USA}
\newcommand{\OskarKlein}{Oskar Klein Centre for Cosmoparticle Physics, Stockholm University, AlbaNova, Stockholm SE-106 91, Sweden}
\newcommand{\Oxford}{The University of Oxford, Oxford OX1 3RH, UK}
\newcommand{\Oxy}{Occidental College, Los Angeles, CA 90041, USA}
\newcommand{\ParisSud}{Universit\'{e} Paris-Sud, LAL, UMR 8607, F-91898 Orsay Cedex, France \& CNRS/IN2P3, F-91405 Orsay, France}
\newcommand{\PI}{Perimeter Institute, Waterloo, Ontario N2L 2Y5, Canada}
\newcommand{\Pitt}{University of Pittsburgh and PITT PACC, Pittsburgh, PA 15260, USA}
\newcommand{\PNNL}{Pacific Northwest National Laboratory ,Richland, WA 99352, USA}
\newcommand{\PNPI}{Petersburg Nuclear Physics Institute, 188300 Gatchina, Russia}
\newcommand{\Port}{Institute of Cosmology \& Gravitation, University of Portsmouth, Dennis Sciama Building, Burnaby Road, Portsmouth PO1 3FX, UK}
\newcommand{\Princeton}{Princeton University, Princeton, NJ 08544, USA}
\newcommand{\PSU}{The Pennsylvania State University, University Park, PA 16802, USA}
\newcommand{\Purdue}{Purdue University, West Lafayette, IN 47907, USA}
\newcommand{\PW}{Participation Worldscope, Sedona, Arizona and Hong Kong, SAR PRC}
\newcommand{\Queens}{Queen's University , K7L 3N6 Kingston, Canada}
\newcommand{\Queensland}{The University of Queensland, School of Mathematics and Physics, QLD 4072, Australia}
\newcommand{\QMUL}{Queen Mary University of London, Mile End Road, London E1 4NS, UK}
\newcommand{\RAL}{Radio Astronomy Laboratory, University of California Berkeley, Berkeley, CA 94720, USA}
\newcommand{\Rice}{Department of Physics \& Astronomy, Rice University, Houston, Texas 77005, USA}
\newcommand{\RIT}{Rochester Institute of Technology}
\newcommand{\RomaS}{Dipartimento di Fisica, Universit\`{a} La Sapienza, P. le A. Moro 2, Roma, Italy}
\newcommand{\RUG}{Kapteyn Astronomical Institute, University of Groningen, P.O. Box 800, 9700 AV Groningen, The Netherlands}
\newcommand{\Rutgers}{Department of Physics and Astronomy, Rutgers, the State University of New Jersey, 136 Frelinghuysen Road, Piscataway, NJ 08854, USA}
\newcommand{\Sanford}{Sanford Underground Research Facility, Lead, SD 57754, USA}
\newcommand{\Sassari}{Universit\`a di Sassari, 07100 Sassari,  Italy}
\newcommand{\SCIPP}{University of California at Santa Cruz, Santa Cruz, CA 95064, USA}
\newcommand{\Sejong}{Department of Physics and Astronomy, Sejong University, Seoul, 143-747, Korea}
\newcommand{\Sheffield}{University of Sheffield, S3 7RH Sheffield, UK}
\newcommand{\SHAO}{Shanghai Astronomical Observatory (SHAO), Nandan Road 80, Shanghai 200030, China}
\newcommand{\Siena}{Siena College, 515 Loudon Road, Loudonville, NY 12211, USA}
\newcommand{\SISSA}{SISSA - International School for Advanced Studies, Via Bonomea 265, 34136 Trieste, Italy}
\newcommand{\SimonFraser}{Department of Physics, Simon Fraser University, Burnaby, British Columbia, Canada V5A 1S6}
\newcommand{\SLAC}{SLAC National Accelerator Laboratory, Menlo Park, CA 94025, USA}
\newcommand{\SMU}{Southern Methodist University, Dallas, TX 75275, USA}
\newcommand{\SNOLAB}{SNOLAB, Lively, ON P3Y 1N2, Canada}
\newcommand{\SoCal}{University of Southern California, CA 90089, USA }
\newcommand{\Stanford}{Stanford University, Stanford, CA 94305, USA}
\newcommand{\StonyBrook}{Stony Brook University, Stony Brook, NY 11794, USA}
\newcommand{\STSCI}{Space Telescope Science Institute, Baltimore, MD 21218, USA}
\newcommand{\SUNYA}{University at Albany SUNY, Albany, NY 12222, USA}
\newcommand{\SussexAstronomy}{Astronomy Centre, School of Mathematical and Physical Sciences, University of Sussex, Brighton BN1 9QH, UK}
\newcommand{\Syracuse}{Syracuse University, Syracuse, NY 13244, USA}
\newcommand{\Tamu}{Texas AandM University, College Station, TX 77843, USA }
\newcommand{\Techsource}{Techsource Incorporated, Los Alamos, NM 87544, USA}
\newcommand{\TelAviv}{Tel-Aviv University,  69978 Tel-Aviv, Israel}
\newcommand{\Temple}{Temple University, Philadelphia, PA 19122, USA}
\newcommand{\TIFR}{Tata Institute of Fundamental Research, Homi Bhabha Road, Mumbai 400005 India}
\newcommand{\Tsinghua}{Department of Physics and Tsinghua Center for Astrophysics, Tsinghua University, Beijing 100084, P R China}
\newcommand{\TUM}{Technical University of Munich,  80333 Munich, Germany}
\newcommand{\UA}{University of Alabama, Tuscaloosa, AL 35487, USA}
\newcommand{\UAS}{Department of Astronomy/Steward Observatory, University of Arizona, Tucson, AZ  85721, USA}
\newcommand{\UAM}{Universidad Aut\'onoma de Madrid, 28049, Madrid, Spain}
\newcommand{\UBC}{University of British Columbia, Vancouver, BC V6T 1Z1, Canada}
\newcommand{\UCB}{Department of Astronomy, University of California Berkeley, Berkeley, CA 94720, USA}
\newcommand{\UCBP}{Department of Physics, University of California Berkeley, Berkeley, CA 94720, USA}
\newcommand{\UCBSSL}{Space Sciences Laboratory, University of California Berkeley, Berkeley, CA 94720, USA}
\newcommand{\UCD}{University of California at Davis, Davis, CA 95616, USA}
\newcommand{\UChicago}{University of Chicago, Chicago, IL 60637, USA}
\newcommand{\UCI}{University of California, Irvine, CA 92697, USA}
\newcommand{\UCLA}{University of California at Los Angeles, Los Angeles,  CA 90095, USA}
\newcommand{\UCL}{University College London, WC1E 6BT London, UK}
\newcommand{\UCR}{University of California at Riverside, Riverside, CA 92521, USA}
\newcommand{\UCSB}{University of California at Santa Barbara, Santa Barbara, CA 93106, USA}
\newcommand{\UCSC}{University of California at Santa Cruz, Santa Cruz, CA 95064, USA}
\newcommand{\UCSD}{University of California San Diego, La Jolla, CA 92093, USA}
\newcommand{\UFL}{University of Florida, Gainesville, FL 32611, USA}
\newcommand{\UFN}{Universit\`a Federico II di Napoli, 80125 Napoli, Italy}
\newcommand{\UGTO}{Divisi\'on de Ciencias e Ingenier\'ias, Universidad de Guanajuato, Le\'on 37150, M\'exico}
\newcommand{\UKY}{University of Kentucky, Lexington, KY 40506, USA}
\newcommand{\UMD}{University of Maryland, College Park, MD 20742, USA}
\newcommand{\UMiami}{University of Miami, Coral Gables, FL 33124, USA}
\newcommand{\UMich}{University of Michigan, Ann Arbor, MI 48109, USA}
\newcommand{\UMN}{University of Minnesota, Minneapolis, MN 55455, USA}
\newcommand{\UnB}{Instituto de F\'{i}sica, Universidade de Bras\'{i}lia, 70919-970, Bras\'{i}lia, DF, Brazil}
\newcommand{\UNC}{University of North Carolina at Chapel Hill, Chapel Hill, NC 27599, USA}
\newcommand{\UNH}{University of New Hampshire, Durham, NH 03824, USA}
\newcommand{\UNIMI}{Dipartimento di Fisica ``Aldo Pontremoli'', Universit\`a{} degli Studi di Milano, via Celoria 16, 20133 Milano, Italy}
\newcommand{\UNIPD}{Dipartimento di Fisica e Astronomia ``G. Galilei'',Universit\`a degli Studi di Padova, via Marzolo 8, I-35131, Padova, Italy}
\newcommand{\UNM}{University of New Mexico, Albuquerque, NM 87131}
\newcommand{\UNV}{University of Nevada, Reno, NV 89557, USA}
\newcommand{\UoM}{Jodrell Bank Center for Astrophysics, School of Physics and Astronomy, University of Manchester, Oxford Road, Manchester, M13 9PL, UK}
\newcommand{\UPenn}{Department of Physics and Astronomy, University of Pennsylvania, Philadelphia, Pennsylvania 19104, USA}
\newcommand{\UR}{Department of Physics and Astronomy, University of Rochester, 500 Joseph C. Wilson Boulevard, Rochester, NY 14627, USA}
\newcommand{\UrbanaC}{Department of Physics, University of Illinois at Urbana-Champaign, Urbana, Illinois 61801, USA}
\newcommand{\USC}{The University of South Carolina, Columbia, SC 29208, USA}
\newcommand{\USD}{The University of South Dakota, Vermillion, SD 57069, USA}
\newcommand{\UTD}{University of Texas at Dallas, Texas 75080, USA}
\newcommand{\Utenn}{The University of Tennessee, Knoxville, TN 37996, USA}
\newcommand{\Utah}{University of Utah, Department of Physics and Astronomy, 115 S 1400 E, Salt Lake City, UT 84112, USA}
\newcommand{\UVA}{University of Virginia, Charlottesville, VA 22903, USA}
\newcommand{\Uvic}{University of Victoria, BC V8P 5C2 Victoria, Canada}
\newcommand{\UWaterloo}{Department of Physics and Astronomy, University of Waterloo, 200 University Ave W, Waterloo, ON N2L 3G1, Canada}
\newcommand{\UWMadison}{Department of Physics, University of Wisconsin - Madison, Madison, WI 53706, USA}
\newcommand{\UW}{University of Washington, Seattle 98195, USA}
\newcommand{\UWC}{Department of Physics \& Astronomy, University of the Western Cape, Cape Town 7535, South Africa}
\newcommand{\Vanderbilt}{Physics \& Astronomy Department, Vanderbilt University, PMB 401807, 2301 Vanderbilt Place, Nashville, TN 37235, USA}
\newcommand{\VSI}{Van Swinderen Institute for Particle Physics and Gravity, University of Groningen, Nijenborgh 4, 9747~AG~Groningen, The~Netherlands}
\newcommand{\VT}{Virginia Tech, Blacksburg, VA 24061, USA}
\newcommand{\VUU}{Virginia Union University, Richmond, Virginia, 23220, USA}
\newcommand{\WCA}{Centre for Astrophysics, University of Waterloo, Waterloo, Ontario N2L 3G1, Canada}
\newcommand{\Weizmann}{Weizmann Institute of Science, 76100 Rehovot, Israel}
\newcommand{\Wellesley}{Wellesley College, Wellesley, MA 02481, USA}
\newcommand{\wiscIce}{University of Wisconsin, Madison, WI 53706, USA}
\newcommand{\WM}{College of William and Mary, Newport News, VA 23606, USA}
\newcommand{\WUSL}{Washington University in St Louis, St. Louis, MO 63130, USA}
\newcommand{\WVU}{CSEE, West Virginia University, Morgantown, WV 26505, USA}
\newcommand{\WVUGWAC}{Center for Gravitational Waves and Cosmology, West Virginia University, Morgantown, WV 26505, USA}
\newcommand{\Wyoming}{Department of Physics and Astronomy, University of Wyoming, Laramie, WY 82071, USA}
\newcommand{\Yale}{Department of Physics, Yale University, New Haven, CT 06520, USA}
\newcommand{\YorkU}{Department of Physics and Astronomy, York University, Toronto, Ontario M3J 1P3, Canada}
\newcommand{\IRAP}{IRAP, Universit\'e de Toulouse, CNRS, CNES, UPS, Toulouse, France}
\newcommand{\AIfA}{Argelander Institute for Astronomy, University of Bonn, Auf dem H\"ugel 71, D-53121 Bonn, Germany}
\newcommand{\Hamburg}{Hamburger Sternwarte, Gojenbergsweg 112, 21029 Hamburg, Germany}
\newcommand{\RIKEN}{Computational Astrophysics Laboratory -- RIKEN, 2-1 Hirosawa, Wako, Saitama, Japan}
\newcommand{\intwopthree}{IN2P3 Computing Center, CNRS, Lyon-Villeurbanne, France}
\newcommand{\LeidenObs}{Leiden Observatory, Leiden University, PO Box 9513, 2300 RA Leiden, The Netherlands}
\newcommand{\LMU}{Ludwig-Maximilians-Universit\"at, Scheinerstr. 1, 81679 Munich, Germany}
\newcommand{\Chicago}{The University of Chicago, Chicago, IL 60637}
\newcommand{\Lagrange}{Laboratoire Lagrange, UMR 7293, Universit\'e de Nice Sophia Antipolis, CNRS, Observatoire de la C\^ote d'Azur, 06304 Nice, France}
\newcommand{\MSU}{Michigan State University, East Lansing, MI 48824-2320, USA}
\newcommand{\Bristol}{University of Bristol, Tyndall Ave, Bristol BS8 1TL, UK}
\newcommand{\CEA}{Service d'Astrophysique, CEA Saclay, Orme des Merisiers, F-91191 Gif-sur-Yvette, France}
\newcommand{\Heidelberg}{Astronomisches Rechen-Institut, Zentrum f\"ur Astronomie der Universit\"at Heidelberg, M\"onchhofstrasse 12-14, D-69120 Heidelberg, Germany}
\newcommand{\ESO}{European Southern Observatory, Garching, Germany}
\newcommand{\IFRU}{IRFU, CEA, Universit\'e Paris-Saclay, 91191, Gif-Sur-Yvette, France}
\newcommand{\UND}{University of North Dakota, Grand Forks, ND 58202}
\renewcommand{\CfA}{Center for Astrophysics $|$ Harvard \& Smithsonian, Cambridge, MA 02138} 
\newcommand{\HDSI}{Harvard Data Science Initiative, Harvard University, Cambridge, MA 02138}
\newcommand{\Bologna}{Universit\`a di Bologna, via Gobetti 93/2, 40129 Bologna, Italy}
\newcommand{\Bonn}{University of Bonn, Bonn, Germany}
\newcommand{\Ljubljana}{University of Ljubljana, Jadranska 19, 1000 Ljubljana, Slovenia}
\newcommand{\Boulder}{University of Colorado, Boulder, CO 80309, USA}
\newcommand{\AMES}{NASA Ames Research Center, Moffett Field, CA 94035, USA}
\newcommand{\UCAS}{University of Chinese Academy of Sciences, Beijing 100049, China}
\newcommand{\Alpes}{Univ.\ Grenoble Alpes, Univ. Savoie  Blanc, CNRS, LAPP, 74000 Annecy, France}
\newcommand{\Nara}{Department of Physics, Nara Women’s University, Kitauoyanishi-machi, Nara, Nara 630-8506, Japan}
\newcommand{\UCDenver}{University of Colorado, Denver, CO 80204, USA}
\newcommand{\Dartmouth}{Department of Physics \& Astronomy, Dartmouth College, Hanover, NH 03755, USA}
\newcommand{\MPA}{Max-Planck-Institut f\"ur Astrophysik, Karl-Schwarzschild-Stra\ss e 1, 85748 Garching, Germany}
\newcommand{\SRIMoscow}{Space Research Institute, Profsoyuznaya 84/32, Moscow 117997, Russia}
\newcommand{\UTA}{University of Texas at Autin, Texas, USA}
\newcommand{\ICfRAR}{International Centre for Radio Astronomy Research, University of Western Australia, 7 Fairway, Crawley, 6009, WA, Australia}
\newcommand{\AlabamaHuntsville}{Department of Physics, University of Alabama in Huntsville, ZP12, Huntsville, AL 35899, USA}
\newcommand{\MSFC}{Marshall Space Flight Center, Huntsville, Alabama, USA}
\newcommand{\Clemson}{Clemson University, Clemson, South Carolina, USA}
\newcommand{\UWBothell}{University of Washington, Bothell, WA 98011, USA}
\newcommand{\UArkansas}{University of Arkansas, Fayetteville, AR 72701}
\newcommand{\IAEMexico}{Instituto de Astronom\'ia sede Ensenada, Universidad Nacional Aut\'onoma de M\'exico, M\'exico}
\newcommand{\GSU}{Georgia State University, Atlanta, GA 30302, USA}
\newcommand{\GIoT}{Georgia Institute of Technology, North Ave NW, Atlanta, GA 30332, USA}
\newcommand{\UoI}{University of Iowa, Iowa City, IA 52242, USA}
\newcommand{\IEECCSIC}{Institute of Space Sciences (IEEC-CSIC) Campus UAB, Carrer de Can Magrans, s/n 08193 Barcelona, Spain}
\newcommand{\CSICUC}{Instituto de Fisica de Cantabria (CSIC-UC), Avenida de los Castros, 39005 Santander, Spain}
\newcommand{\BMCC}{Borough of Manhattan Community College, The City University of New York, 199 Chambers Street, New York, NY 10007}
\newcommand{\NAChile}{N\'{u}cleo de Astronom\'{\i}a, Facultad de Ingenier\'{\i}a y Ciencias, Universidad Diego Portales, Santiago, Chile}
\newcommand{\Southampton}{Department of Physics \& Astronomy, Faculty of Physical Sciences and Engineering, University of Southampton, Southampton, SO17 1BJ, UK}
\newcommand{\Swinburne}{Swinburne University of Technology, John St, Hawthorn VIC 3122, Australia}
\newcommand{\JMU}{James Madison University, 800 S Main St, Harrisonburg, VA 22807, USA}
\newcommand{\Montreal}{Universit\'e de Montr\'eal, Montr\'eal, Quebec, Canada}
\newcommand{\CdA}{Centro de Astrobiolog\'ia (CAB, CSIC-INTA), 28850 Torrej\'on de Ardoz, Spain}
\newcommand{\INAFRome}{INAF Osservatorio Astronomico di Roma, Via Frascati 33, Monte Porzio Catone, Italy}
\newcommand{\Kennesaw}{Kennesaw State University, Kennesaw, Georgia, USA}
\newcommand{\Leicester}{University of Leicester, University Rd, Leicester LE1 7RH, UK}
\newcommand{\KIfASR}{MIT Kavli Institute for Astrophysics and Space Research. Massachusetts Institute of Technology. 77 Massachusetts Avenue, 37-241. Cambridge, MA 02139, USA}
\newcommand{\IRyA}{Instituto de Radioastronom\'ia y Astrof\'isica, Universidad Nacional Aut\'onoma de M\'exico, M\'exico}
\newcommand{\CalStateNorthridge}{California State University Northridge, 18111 Nordhoff St, Northridge, CA 91330, USA}
\newcommand{\Calgary}{University of Calgary, 2500 University Drive NW Calgary, AB T2N 1N4 Canada}
\newcommand{\FIoT}{Florida Institute of Technology, 150 W University Blvd, Melbourne, FL 32901, USA}
\newcommand{\USTChina}{University of Science and Technology of China, 1129 Huizhou Ave, Baohe Qu, Hefei Shi, Anhui Sheng, China}
\newcommand{\Eotvos}{MTA-E\"otv\"os Lor\'and University, Hungary/Masaryk University, Czech Republic}
\newcommand{\PUCChile}{Pontifical Catholic University of Chile, Instituto de Astrof\'isica, Santiago, Chile}
\newcommand{\NDLouaize}{Dept of Physics \& Astronomy, Notre Dame University–Louaize, Lebanon}
\newcommand{\IoATaiwan}{Institute of Astronomy, National Tsing Hua University, Taiwan}
\newcommand{\SSI}{Spectral Sciences, Inc.}
\newcommand{\TTU}{Texas Tech University, 2500 Broadway, Lubbock, TX 79409, USA}
\newcommand{\CASSACA}{Chinese Academy of Sciences South American Center for Astronomy, China}
\newcommand{\OpenU}{The Open University, Milton Keynes, UK}

\textbf{Principal Authors:}\\
Francesca Civano$^{1}$ (E-mail: fcivano@cfa.harvard.edu; Tel: +1 617-792-3190), Nico Cappelluti$^{2}$, Ryan Hickox$^{3}$, Rebecca Canning$^{4,5}$ 

\noindent 
\textbf{Co-authors:} \\
James Aird $^{6}$,   
Marco Ajello $^{7}$,   
Steve Allen $^{4,5}$,   
Eduardo Ba\~nados $^{8}$,   
Laura Blecha $^{9}$,   
William N. Brandt $^{10}$,  
Marcella Brusa $^{11}$,   
Francisco Carrera $^{12}$,   
Massimo Cappi $^{13}$,   
Andrea Comastri $^{13}$,   
Klaus Dolag $^{14,15}$,   
Megan Donahue $^{16}$,   
Martin Elvis $^{1}$,   
Giuseppina Fabbiano $^{1}$,   
Francesca Fornasini $^{1}$,   
Poshak Gandhi $^{17}$,   
Antonis Georgakakis $^{18}$,   
Kelly Holley-Bockelmann $^{19}$,   
Anton Koekemoer $^{20}$,   
Andrew Goulding $^{21}$,   
Mackenzie Jones $^{1}$,   
Sibasish Laha $^{22}$,   
Stephanie LaMassa $^{20}$,   
Giorgio Lanzuisi $^{13}$,   
Lauranne Lanz $^{3}$,   
Adam Mantz $^{4,5}$,   
Stefano Marchesi $^{7}$,   
Mar Mezcua $^{23}$,   
Beatriz Mingo $^{24}$,   
Kirpal Nandra $^{18}$,   
Daniel Stern $^{25,26}$,   
Doug Swartz $^{27}$,   
Grant Tremblay $^{1}$,   
Panayiotis Tzanavaris $^{28}$,   
Alexey Vikhlinin $^{1}$,   
Fabio Vito $^{29,30}$,   
Belinda Wilkes $^{1}$\\\\

\noindent
\textbf{Abstract:}\\
\noindent The discoveries made over the past 20 years by \chandra\ and \xmm\ surveys in conjunction with multiwavelength imaging and spectroscopic data available in the same fields have significantly changed the view of the supermassive black hole (SMBH) and galaxy connection. These discoveries have opened up several exciting questions that are beyond the capabilities of current X-ray telescopes and will need to be addressed by observatories in the next two decades. As new observatories peer into the early Universe, we will begin to understand the physics and demographics of SMBH infancy (at $z>6$) and investigate the influence of their accretion on the formation of the first galaxies ($\S$ 2.1). We will also be able to understand the accretion and evolution over the cosmic history (at $z\sim$1--6) of the full population of black holes in galaxies, including low accretion rate, heavily obscured AGNs at luminosities beyond the reach of current X-ray surveys ($\S$2.2 and $\S$2.3), enabling us to resolve the connection between SMBH growth and their environment.\\

\noindent 
\textbf{Endorsers:} \\
Scott Anderson $^{31}$,   
Vallia Antoniou $^{1,32}$,   
Manuel Aravena $^{33}$,   
Ashraf Ayubinia $^{34}$,   
Mitchell Begelman $^{35}$,   
Arash Bodaghee $^{36}$,   
Tamara Bogdanovi\'c $^{37}$,   
Rozenn Boissay-Malaquin $^{38}$,   
Angela Bongiorno $^{39}$,   
Peter Boorman $^{17}$,   
Stefano Borgani $^{40}$
% ...
\\

 \vspace{-0.2cm}
\noindent 
$^{1}$ \CfA\\
$^{2}$ \UMiami\\
$^{3}$ \Dartmouth\\
$^{4}$ \Stanford\\
$^{5}$ \KIPAC\\
$^{6}$ \Leicester\\
$^{7}$ \Clemson\\
$^{8}$ \MPIA\\
$^{9}$ \UFL\\
$^{10}$ \PSU\\
$^{11}$ \Bologna\\
$^{12}$ \CSICUC\\
$^{13}$ \INAFOAS\\
$^{14}$ \LMU\\
$^{15}$ \MPA\\
$^{16}$ \MSU\\
$^{17}$ \Southampton\\
$^{18}$ \MPE\\
$^{19}$ \Vanderbilt\\
$^{20}$ \STSCI\\
$^{21}$ \Princeton\\
$^{22}$ \UCSD\\
$^{23}$ \IEECCSIC\\
$^{24}$ \OpenU\\
$^{25}$ \Caltech\\
$^{26}$ \JPL\\
$^{27}$ \MSFC\\
$^{28}$ \GSFC\\
$^{29}$ \PUCChile\\
$^{30}$ \CASSACA\\
$^{31}$ \UW\\
$^{32}$ \TTU\\
$^{33}$ \NAChile\\
$^{34}$ \USTChina\\
$^{35}$ \Boulder\\
$^{36}$ \GSU\\
$^{37}$ \GIoT\\
$^{38}$ \KIfASR\\
$^{39}$ \INAFRome\\
$^{40}$ \INAFOATs\\
% ...


\justify

\section{Introduction}
\vspace{- 0.35 cm}
The past two decades have revealed strong links between the cosmic evolution of supermassive black holes (SMBHs), their host galaxies, and dark matter halos.
In the local Universe, SMBH masses are tightly correlated with their host-galaxy bulge luminosity and stellar velocity dispersion (e.g., Kormendy \& Ho 2013), while at higher redshifts, the global density of star formation and SMBH growth follow a common trend, rising from the epoch of reionization to $z\sim$ 2--3, then steeply declining to the current epoch. 
Feedback from growing SMBHs as active galactic nuclei (AGNs; see Laha et al. 2019, Tombesi et al. 2019) is critical for reproducing the observed populations of galaxies and their hot halos in models (e.g., McAlpine et al. 2018, Weinberger et al. 2018). Despite this progress, the nature of SMBH fueling and the interplay between SMBHs and their hosts remain unclear, particularly before the peak epoch of structure formation.

To uncover the co-evolution of SMBHs and galaxies, we must probe the full scope of cosmic time to detect and characterize large samples of AGNs and their hosts, including sources that are heavily obscured by gas and dust (e.g., Hickox \& Alexander 2018). While AGNs can be identified and studied across a range of wavelengths (e.g., Padovani et al. 2017), a particularly powerful waveband for AGN selection is in the X-rays. At X-ray energies, the contamination from non-nuclear emission, mainly due to star formation processes, can be far less significant than in broad-band optical and IR observations. This white paper will focus on X-ray observations.

\begin{figure}
\begin{center}
\includegraphics[width=0.95\textwidth]{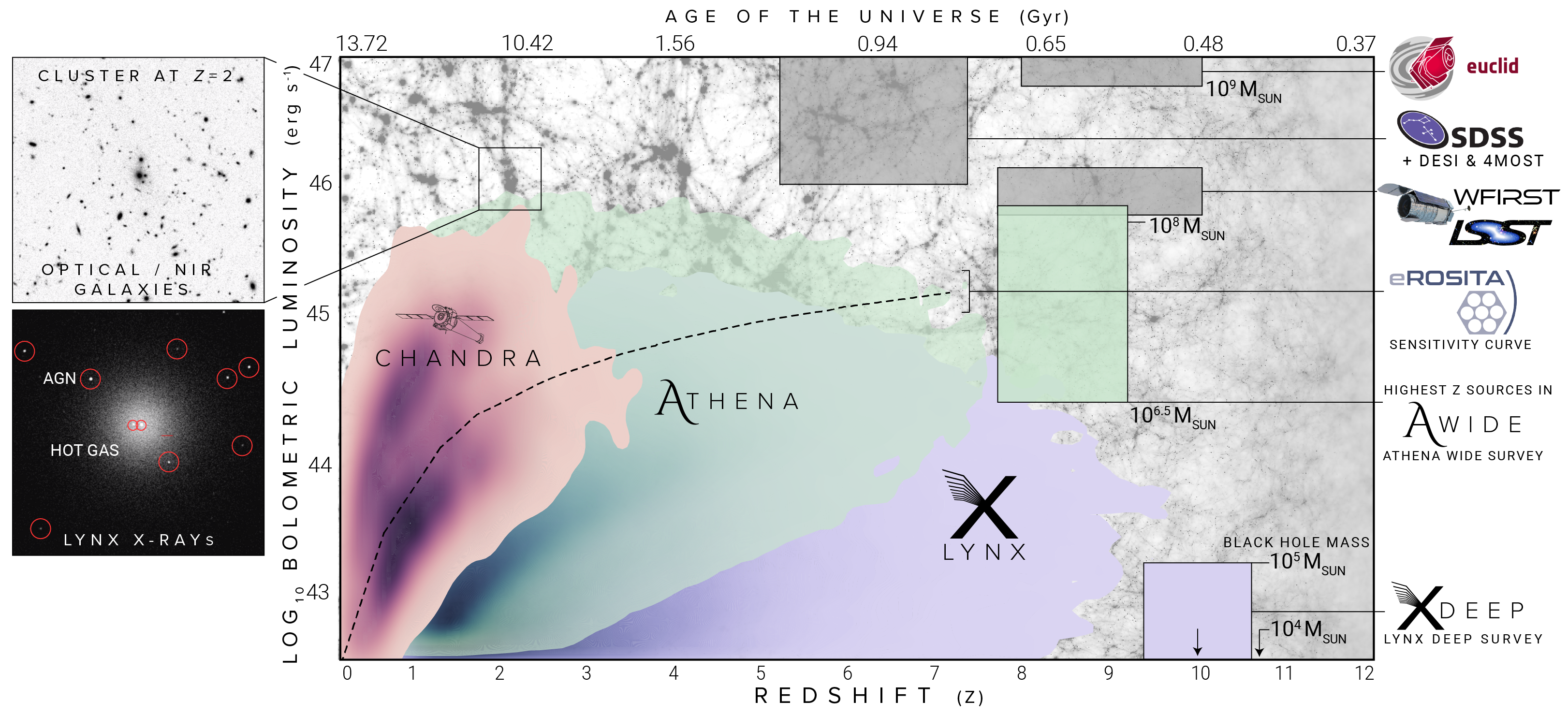}
\end{center}
\vspace{-0.8cm}
\caption{\footnotesize{Redshift and bolometric luminosity (with corresponding SMBH masses for Eddington-limited accretion) distributions to be probed by future X-ray facilities ({\em Athena} and {\em Lynx}), compared with current \chandra\ surveys and the {\it eROSITA} limit at the ecliptic poles. In this figure, the total time allocated to surveys for future mission is $\sim$ 25 Ms. The ranges covered by high-redshift optical and NIR wide surveys are reported (gray boxes). The left inset shows how high-resolution observatories will resolve AGNs in groups and cluster of galaxies at $z\sim$~2 (see Mantz et al. 2019). \vspace{-0.5cm}}}
\label{lxz}
\end{figure}

Over the past 20 years, \chandra\ and \xmm\ X-ray surveys have successfully collected a clean and largely unbiased sample of AGNs ($\sim$60,000) up to $z=5$ (e.g., Brusa et al. 2010, Nandra et al. 2015, LaMassa et al. 2016, Civano et al. 2016, Luo et al. 2017, Chen et al. 2018, Masini et al. in prep.). These samples cover a broad range of luminosities (down to $L_X\sim 10^{44}$ erg s$^{-1}$ at $z=5$ and $L_X=10^{39}$ erg s$^{-1}$ at $z=0.5$; see Fig. \ref{lxz}), corresponding to BH masses of $10^6-10^9$ $M_\odot$, and include large numbers of both unobscured and obscured AGNs. 
In concert with multiwavelength data, X-ray observations have made a number of key discoveries: (1) massive ($>$10$^9$ M$\odot$) accreting SMBHs already existed at $z\sim$~7.5, challenging seed formation models; (2) AGN number densities peak at $z\sim$~2--3, with luminous sources peaking earlier in cosmic time and with most of the accretion being obscured; (3) correlations exist between SMBH accretion and galaxy properties, particularly with stellar/dark matter halo mass and (perhaps indirectly) star formation, although due to AGN stochasticity these can only be revealed in statistical samples. 
 
These discoveries have opened up several exciting questions that are beyond the capabilities of current facilities and will need to be addressed by new observatories into the next two decades: \vspace{-0.25cm}
\begin{itemize}
\setlength\itemsep{0.0001ex}
\item {\bf How did the first SMBHs form and grow so rapidly in the early Universe}? \vspace{-0.1cm}
\item{ \bf What is the complete census of growing SMBHs from cosmic dawn ($\bm{z\sim}$6) to the peak formation epoch ($\bm{z\sim}$2) and beyond?}\vspace{-0.1cm}
\item{\bf How does SMBH accretion influence the growth of galaxies and large-scale structures?}\vspace{-0.1cm}
\end{itemize}
In this white paper, we will discuss these questions and also highlight how some future missions, such as ESA’s {\em Athena} (Nandra et al. 2013), and proposed concept missions, such as {\em Lynx} (The Lynx team 2018) and {\em AXIS} (Mushotzky et al. 2018), will help us address them. 

\vspace{-0.6cm}
\section{Understanding the build up of SMBHs and galaxies} \vspace{-0.3cm}
\subsection{Black holes and galaxies at cosmic dawn}\vspace{-0.2cm}
The ultimate origin of SMBHs is still unknown and remains one of the most pressing open questions in astrophysics. Optical and NIR surveys have revealed the existence of SMBHs with masses of 10$^9$ M$_{\odot}$ up to $z\sim$~7.5 when the Universe was only 600 Myr old (e.g., Mortlock et al. 2011, Ba\~nados et al. 2018). This discovery lead to the proposal of several BH seed formation models, including: (1) direct collapse of early pristine massive clouds (Begelman et al. 2006); (2) Pop III star remnants (Bromm \& Loeb 2006); (3) coalescence of stellar mass BHs in star clusters (e.g., Devecchi \& Volonteri 2009); and (4) primordial BHs formed during inflation (Hawking 1971).
					
A complete understanding of SMBH origins and their contributions to reionization demands both the census of massive, luminous AGNs at $z > 6$ as well as direct measurements of the lower-mass seeds, forming and growing at $z=$~10--15. The first can be observed by surveying wide areas (tens of deg$^2$) at the equivalent of the faintest \chandra\ flux ($\sim 10^{-17}$ erg s$^{-1}$ cm$^{-2}$), e.g. with the proposed {\em Athena-WFI} survey strategy (as in Fig. \ref{lxz}; Aird et al. 2013). The second requires extreme X-ray detection limits ($10^{-19}$ erg s$^{-1}$ cm$^{-2}$), implying an observatory with both large effective area and sub-arcsecond angular resolution, to avoid source confusion, such as e.g. {\em Lynx} and {\em AXIS}. 

\begin{SCfigure}
\vspace{1cm}
\includegraphics[width=0.55\textwidth]{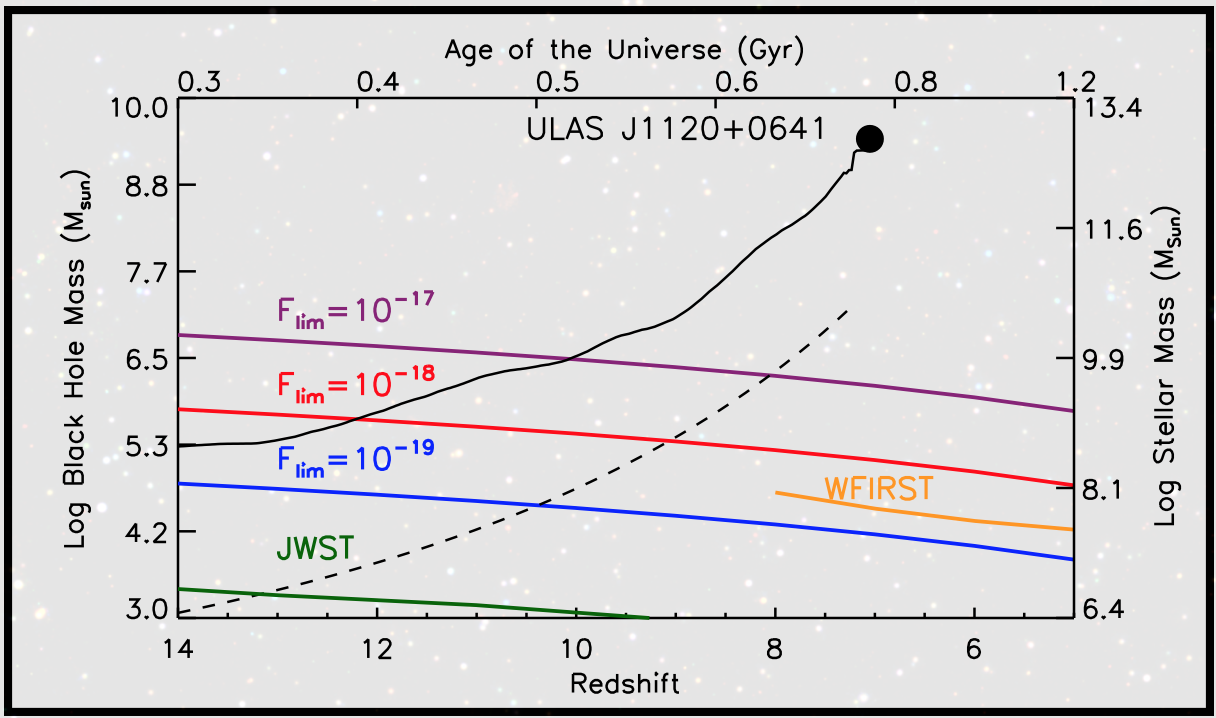}
\vspace{1cm}
\caption{\footnotesize{Simulated growth of a massive BH seed (10$^5$ M$_{\odot}$ at z=20, solid) up to the the size of $ULAS 1120+0641$ and of a low-mass seed (10$^2$ M$_\odot$ at z=20, dashed) using the prescriptions in Pacucci et al. (2017) and continuous Eddington limited growth, respectively. These are compared with 0.5-2 keV limiting sensitivities for {\it Athena} (magenta), AXIS (red) and {\it Lynx} (blue). The BH mass limits (derived from M$_{\star}$; Song et al. 2016) detectable by {\em JWST} (green) and {\em WFIRST} (orange) observations are reported. X-ray observatories will detect the low mass BHs in the first galaxies detected by {\em JWST} and {\em WFIRST}. }}
\vspace{-2cm}
\label{BH_seed}
\end{SCfigure}

A major challenge will be obtaining rest-frame optical counterparts for these X-ray sources, since these will be extremely faint at all wavelengths (e.g., Volonteri et al. 2017). In Fig. \ref{BH_seed}, we compare the limiting BH masses detectable in X-ray surveys with a flux limit range from $10^{-17}$ erg s$^{-1}$ cm$^{-2}$, the {\em Athena} flux limit, to $10^{-19}$ erg s$^{-1}$ cm$^{-2}$, the ultimate depth reachable with {\em Lynx}. {\em AXIS}, with comparable angular resolution to {\em Lynx} but a smaller effective area, will reach a depth between these two limits.
Fig.~\ref{BH_seed} shows that the combination of {\em Athena} and {\em WFIRST} will provide X-ray and NIR information for growing seeds at $z\sim$~5--8 and {\em Lynx} (or {\em AXIS}) will peer into the early Universe to $z\geq$~10 and at lower mass, in concert with {\em JWST}. 

While predictions for the number density of early BH seeds are still uncertain (Haiman et al. 2019), it is clear that combining a deep X-ray survey with sub-arcsecond positional accuracy with {\em JWST} and {\em WFIRST} data will return a sample of tens to thousands of X-ray selected SMBH seeds with confirmed counterparts (Fig. \ref{lxz}). With these samples, it will be possible to probe, for the first time, the luminosity function down to $L_X = 10^{43}$ erg s$^{-1}$ at $z\sim$~15, corresponding to $10^{4-5}$~M$_\odot$ BHs. Progress will also be made by studying the fluctuations of the X-ray background and its counterparts (see e.g., Kashlinsky et al. 2018, Cappelluti et al. 2013, 2017), but the strongest constraints will come from direct detections of lower-mass seed BHs in the ``first stars'' epoch. 

\vspace{-0.5cm}
\subsection{The full population of black holes from cosmic dawn to cosmic noon and the connection with the host galaxy} \vspace{-0.2cm}
Beyond the origin of SMBHs, studying their growth from ``cosmic dawn'' at $z >$6 to ``cosmic noon'' at $z\sim$1--2 is essential for understanding their co-evolution with galaxies, in the epochs when the bulk of the Universe’s stars were formed. A recent breakthrough in our understanding of AGNs shows that they are not like light bulbs but instead ``flicker'' across a wide range of Eddington ratios on relatively short time scales ($<10^{5-6}$ years; Hickox et al. 2014). Thus, to understand the full AGN population, the full dynamic range in accretion rates (and thus the full luminosity function) needs to be probed across all of cosmic time. This approach has been critical for understanding the connection between BH and galaxy growth, the nature of AGN obscuration and the evolution of AGNs in large-scale structures (e.g., Chen et al. 2013, Yang et al. 2017, Powell et al. 2018).

Observations across a wide range of wavelengths are essential for this work. Optical and IR imaging and spectroscopy can detect luminous AGNs outshining their host galaxies and can characterize the hosts of weak or obscured AGNs, while far-IR--mm observations are critical for constraining star formation and dust masses. However, due to high penetrating power and contrast with host galaxy light, X-ray observations are required for probing the full range of SMBH masses and Eddington ratios. The most recent measurements of the luminosity function probe the X-ray luminosity range $10^{42}$--$10^{43}$ erg s$^{-1}$ to $z=$~2 and $10^{45}$--$10^{46}$ erg s$^{-1}$ to $z=$~5 (e.g., Ueda et al. 2014, Miyaji et al. 2015, Aird et al. 2015, Tasnim Ananna et al. 2019). In the next few years, {\it eROSITA} (Merloni et al. 2012) will constrain the bright end of the luminosity function ($> 10^{45}$ erg s$^{-1}$) with $\sim 10^6$ AGNs over a broad redshift range (Fig. \ref{lxz}).

Furthermore, it is critical to probe deeper, below the ``knee'' of the luminosity function at the earliest epochs ($z >$ 3), requiring large numbers of very faint sources (many thousands compared to the few hundreds available now; see Marchesi et al. 2016, Vito et al. 2018). The first step will be to cover large areas down to the flux limits of the current deepest X-ray survey (the \chandra\ Deep Field-South). This will be feasible with a large effective area telescope (e.g., {\em Athena}), identifying tens of thousands of distant AGNs over tens of deg$^{2}$, to a flux limit currently available only in pencil-beam (0.1~deg$^2$) fields (Fig.~\ref{lxz}). The next step will be to probe to significantly fainter fluxes ($\sim$ 10$^{-19}$ erg s$^{-1}$ cm$^{-2}$), constraining the X-ray luminosity and Eddington ratio distributions at least an order of magnitude fainter. Such deep flux limits require {\em both} high effective area {\em and} angular resolution (e.g., as proposed for {\em Lynx}; Fig.~\ref{lxz}). As a complementary approach, it will be possible to use X-ray stacking techniques with high-resolution X-ray data (e.g., Vito et al. 2016, Fornasini et al. 2018) to push to even fainter average fluxes ($\sim$ 10$^{-20}$ erg s$^{-1}$ cm$^{-2}$).

Counterpart identifications will be readily available by combining data from next generation space facilities ({\em JWST}, {\em WFIRST}, {\em Euclid}) as well as ground based 30m telescopes (ELT, GMT, TMT), allowing us to explore the distribution of AGN accretion rates as a function of key host properties (i.e., star formation rate, stellar mass, morphology), and perform statistical analyses of clustering to probe the connection between AGN and their host environment (e.g., Mendez et al. 2016, Allevato et al. 2016, Plionis et al. 2018, Yang et al. 2018).


		
Finally, reaching the very faintest luminosity limits, we will probe down into the galaxy population with emission dominated by X-ray binaries and hot gas (L$_X<10^{41}$ erg s$^{-1}$) as well as group and cluster-sized halos at $z\sim$~2. High spatial resolution observatories can distinguish point sources above this diffuse background, enabling simultaneous measurements of AGN properties and a {\em direct} measure of how AGNs populate large-scale structures (see inset in Fig. \ref{lxz} and Mantz et al. 2019). Following up halos detected with CMBs-4 and {\it Athena}, {\it Lynx} will build on the legacy of \chandra\ and will map the triggering of SMBHs in the build-up of large scale structure out to $z\sim$~4. In lower-redshift systems, high-resolution X-ray imaging and spectrocopy will directly probe the effects of AGN feedback on the hot gas atmospheres in their host galaxies.

\vspace{-0.5cm}
\subsection{Uncovering the ``hidden'' black hole population} \vspace{-0.2cm}
A significant number of AGNs are ``hidden'' by dust that only X-ray light can penetrate. The majority of this population are obscured by a dusty ``torus'' in close proximity to the SMBH, but dust at larger radii in the host galaxy can also shield the AGN from view. 

\begin{figure}
\begin{center}
\includegraphics[width=0.76\textwidth]{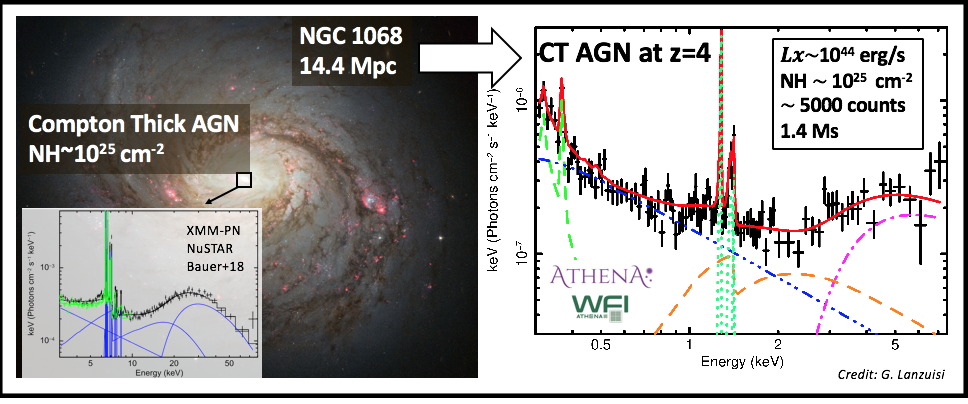}
\end{center}
\vspace{-0.75cm}
\caption{\footnotesize{{\it Athena} will enable detailed spectral analysis of even the most obscured sources. The simulated X-ray spectrum of a $z=$4 AGN with $N_H \sim 10^{25}$ cm$^{-2}$ as seen by {\it Athena} (right) is compared with the spectrum of the well known NGC~1068 (left), showing the same obscuring properties in the local Universe. }}
\vspace{-0.6cm}
\label{CT_AGN}
\end{figure}

At low-redshifts, around a third of the AGN population are known to be obscured even to hard X-rays (e.g., Risaliti et al. 1999). Dedicated observing campaigns have been focusing on some of the extreme cases (e.g., NGC 1068 with $N_H \sim 10^{25}$ cm$^{-2}$; e.g. Bauer et al. 2015; see Fig.\ref{CT_AGN}). 
However, little is known about these sources at higher redshifts: only a few 100 candidates have been detected thus far (e.g., Brightman et al. 2014, Lanzuisi et al. 2018) and the evolution of the obscured fraction is poorly constrained (e.g., Liu et al. 2017, Zappacosta et al. 2018). Pioneering work by {\it NuSTAR}, operating above 20 keV, is now uncovering some of these concealed sources (Civano et al. 2015, Lansbury et al. 2017, Masini et al. 2018) but still with low numbers of photons. Theoretical models predict phases of high obscuration during gas-rich galaxy mergers when feedback from AGN may play an important role in regulating galaxy growth (e.g., Hopkins et al. 2005, Blecha et al. 2018). It is thus critical to uncover and properly model these hidden AGNs, even at the highest redshifts, to understand the connection between the obscured growth of SMBHs, their host galaxies, and the connection with mergers (which can in turn be probed directly with future space-based gravitational wave observatories, e.g. {\em LISA}). 

Reprocessed IR emission offers one way of detecting such AGNs (e.g., Stern et al. 2005, Alexander et al. 2008, Gandhi et al. 2009, Goulding et al. 2011). Unfortunately, this suffers from both contamination and dilution by the host galaxy. High sensitivity and spatial resolution NIR observations with {\em JWST} and {\em WFIRST} will begin to uncover these low-to-medium redshift AGNs buried within their host emission. We can directly observe reprocessed AGN emission in the high-energy X-rays, via the `Compton hump' at 20 -- 40 keV and a ubiquitous iron emission line at 6.4 keV. At z$>$2, the Compton hump is redshifted to low-energy X-rays, making the complex spectrum of these sources accessible by soft X-ray telescopes (Fig. \ref{CT_AGN}). 

Probe missions in the soft and hard X-rays (e.g., {\em STROBE-X} and {\em HEX-P}, see Hickox et al. 2019) will inform us about the most luminous obscured AGNs out to $\sim$2. Only the leap in sensitivity over {\em Chandra} and {\em XMM-Newton} combined with a wide field of view, e.g. such as with {\it Athena}, will push the census of heavily obscured AGNs beyond the peak epoch of activity out to $z=$4. We will be able to perform unprecedented spectral modeling, even at high redshifts, uncovering the physics of extremely obscured AGN, at levels currently available only for local sources (Fig. \ref{CT_AGN}; e.g., Georgakakis et al. 2013).  At even higher redshifts, the combination of effective area and spatial resolution will enable a mission such as {\it Lynx} to both uncover the least luminous of these hidden SMBHs and move toward a complete census of SMBH growth across cosmic time.

\vspace{-0.5cm}
\section{Our vision for an extragalactic survey program into the 2030s} \vspace{-0.2cm}
This white paper outlines a very exciting range of science that can be completed with an extragalactic survey program in the next two decades, demanding a substantial leap forward in observing capabilities across a wide range of wavelengths. A full picture of SMBH growth and galaxy evolution requires a tiered, coordinated survey of all components: {\em starlight} in the rest-frame visible and NIR (planned or proposed facilities include {\em WFIRST}, {\em JWST}, LSST, {\em HabEx}, {\em LUVOIR}), {\em gas and dust} in the mid- and far-IR, microwaves, radio as well as sensitive optical/NIR spectroscopy and X-ray imaging ({\em JWST}, {\em OST}, ALMA, ngVLA, SKA, CMB-S4, 30m class telescopes, {\em XRISM} and {\em Athena}), and the signatures of {\em SMBH growth}, most powerfully probed in X-rays, IR, radio, and GW ({\em Athena}, {\em Lynx}, {\em AXIS}, {\em JWST}, {\em OST}, SKA, {\em LISA}). X-ray observations will produce the biggest breakthroughs in uncovering the complete AGN population. In the next 5 years, {\it Chandra Source Catalog 2} and {\em eROSITA} AGNs will be spectroscopically characterized by SDSS-V, 4MOST and DESI. Afterwards, {\em Athena} will provide a huge increase in effective area and field of view, allowing us to reach the current faintest X-ray flux levels in the deepest {\em Chandra} observations but over a vastly larger area (tens of deg$^{2}$). {\em Lynx} would then provide another enormous leap forward with exquisite angular resolution, allowing us to push more than an order of magnitude fainter and directly resolve and study the first BHs in the early Universe. ({\em AXIS} would make a comparable leap in resolution but with a more modest yield in high-redshift AGNs.) We urge the Astro2020 Decadal Survey Committee to consider the compelling astrophysical questions about the cosmic evolution of SMBH growth outlined here and endorse a plan to address them in the next two decades.



\pagebreak
\noindent 
\textbf{References}\\
$\bullet$ Aird, J., Comastri, A., Brusa, M., et al.\ 2013, arXiv:1306.2325 \\
$\bullet$ Aird, J., Coil, A.~L., Georgakakis, A., et al.\ 2015, MNRAS, 451, 1892 \\
$\bullet$ Alexander, D.~M., Chary, R.-R., Pope, A., et al.\ 2008, ApJ, 687, 835 \\
$\bullet$ Allevato, V., Civano, F., Finoguenov, A., et al.\ 2016, ApJ, 832, 70 \\
$\bullet$ Ba{\~n}ados, E., Venemans, B.~P., Mazzucchelli, C., et al.\ 2018, Nature, 553, 473 \\
$\bullet$ Bauer, F.~E., Ar{\'e}valo, P., Walton, D.~J., et al.\ 2015, ApJ, 812, 116 \\
$\bullet$ Begelman, M.~C., Volonteri, M., \& Rees, M.~J.\ 2006, MNRAS, 370, 289 \\
$\bullet$ Blecha, L., Snyder, G.~F., Satyapal, S., \& Ellison, S.~L.\ 2018, MNRAS, 478, 3056 \\
$\bullet$ Brightman, M., Nandra, K., Salvato, M., et al.\ 2014, MNRAS, 443, 1999 \\
$\bullet$ Bromm, V., \& Loeb, A.\ 2006, ApJ, 642, 382 \\
$\bullet$ Brusa, M., Civano, F., Comastri, A., et al.\ 2010, ApJ, 716, 348\\
$\bullet$ Cappelluti, N., Arendt, R., Kashlinsky, A., et al.\ 2017, ApJL, 847, L11 \\
$\bullet$ Cappelluti, N., Kashlinsky, A., Arendt, R.~G., et al.\ 2013, ApJ, 769, 68 \\
$\bullet$ Chen, C.-T.~J., Hickox, R.~C., Alberts, S., et al.\ 2013, ApJ, 773, 3 \\
$\bullet$ Chen, C.-T.~J., Brandt, W.~N., Luo, B., et al.\ 2018, MNRAS, 478, 2132 \\
$\bullet$ Civano, F., Hickox, R.~C., Puccetti, S., et al.\ 2015, ApJ, 808, 185 \\
$\bullet$ Civano, F., Marchesi, S., Comastri, A., et al.\ 2016, ApJ, 819, 62 \\
$\bullet$ Devecchi, B., \& Volonteri, M.\ 2009, ApJ, 694, 302 \\
$\bullet$ Fornasini, F.~M., Civano, F., Fabbiano, G., et al.\ 2018, ApJ, 865, 43 \\
$\bullet$ Gandhi, P., Horst, H., Smette, A., et al.\ 2009, A\&A, 502, 457 \\
$\bullet$ Georgakakis, A., Carrera, F., Lanzuisi, G., et al.\ 2013, arXiv:1306.2328 \\
$\bullet$ Goulding, A.~D., Alexander, D.~M., Mullaney, J.~R., et al.\ 2011, MNRAS, 411, 1231 \\
$\bullet$ Haiman Z.  et al. 2019, {\it Astro2020 Science White Paper: ``Electromagnetic Window into the Dawn of Black Holes''}\\
$\bullet$ Hawking, S.\ 1971, MNRAS, 152, 75 \\
$\bullet$ Hickox, R.~C., Mullaney, J.~R., Alexander, D.~M., et al.\ 2014, ApJ, 782, 9 \\
$\bullet$ Hickox, R.~C., \& Alexander, D.~M.\ 2018, ARAA, 56, 625 \\
$\bullet$ Hickox, R.~C., Civano F., et al. 2019, {\it Astro2020 Science White Paper: ``Resolving the cosmic X-ray background with a next-generation high-energy X-ray observatory''}\\
$\bullet$ Hopkins, P.~F., Hernquist, L., Martini, P., et al.\ 2005, ApJl, 625, L71 \\
$\bullet$ Kashlinsky, A., Arendt, R.~G., Atrio-Barandela, F., et al.\ 2018, Reviews of Modern Physics, 90, 025006 \\
$\bullet$ Kormendy, J., \& Ho, L. C. 2013, ARAA, 51, 511\\
$\bullet$ Laha, S., et al. 2019, {\it Astro2020 Science White Paper: ``The physics and astrophysics of X-ray outflows from Active Galactic Nuclei.''}\\
$\bullet$ LaMassa, S.~M., Urry, C.~M., Cappelluti, N., et al.\ 2016, ApJ, 817, 172 \\
$\bullet$ Liu, T., Tozzi, P., Wang, J.-X., et al.\ 2017, ApJs, 232, 8 \\
$\bullet$ Lansbury, G.~B., Alexander, D.~M., Aird, J., et al.\ 2017, ApJ, 846, 20 \\
$\bullet$ Lanzuisi, G., Civano, F., Marchesi, S., et al.\ 2018, MNRAS, 480, 2578 \\
$\bullet$ Luo, B., Brandt, W.~N., Xue, Y.~Q., et al.\ 2017, ApJS, 228, 2 \\
$\bullet$ Mantz, A., et al. 2019, {\it Astro2020 Science White Paper: ``The Future Landscape of High-Redshift Galaxy Cluster Science'' }\\
$\bullet$ Marchesi, S., Civano, F., Salvato, M., et al.\ 2016, ApJ, 827, 150 \\
$\bullet$ Masini, A., Comastri, A., Civano, F., et al.\ 2018, ApJ, 867, 162 \\
$\bullet$ McAlpine, S., Bower, R.~G., Rosario, D.~J., et al.\ 2018, MNRAS, 481, 3118 \\
$\bullet$ Mendez, A.~J., Coil, A.~L., Aird, J., et al.\ 2016, ApJ, 821, 55 \\
$\bullet$ Merloni, A., Predehl, P., Becker, W., et al.\ 2012, arXiv:1209.3114 \\
$\bullet$ Miyaji, T., Hasinger, G., Salvato, M., et al.\ 2015, ApJ, 804, 104 \\
$\bullet$ Mortlock, D.~J., Warren, S.~J., Venemans, B.~P., et al.\ 2011, Nature, 474, 616 \\
$\bullet$ Mushotzky, R.\ 2018, Space Telescopes and Instrumentation 2018: Ultraviolet to Gamma Ray, 10699, 1069929 \\
$\bullet$ Nandra, K., Barret, D., Barcons, X., et al.\ 2013, arXiv:1306.2307 \\
$\bullet$ Nandra, K., Laird, E.~S., Aird, J.~A., et al.\ 2015, ApJS, 220, 10 \\
$\bullet$ Pacucci, F., Natarajan, P., Volonteri, M., Cappelluti, N., \& Urry, C.~M.\ 2017, ApJL, 850, L42 \\
$\bullet$ Padovani, P., Alexander, D.~M., Assef, R.~J., et al.\ 2017, A\&A Review, 25, 2 \\
$\bullet$ Plionis, M., Koutoulidis, L., Koulouridis, E., et al.\ 2018, A\&A, 620, A17 \\
$\bullet$ Powell, M. C., Cappelluti, N., Urry, C. M., et al. 2018, ApJ, 858, 110\\
$\bullet$ Risaliti, G., Maiolino, R., \& Salvati, M.\ 1999, ApJ, 522, 157 \\
$\bullet$ Song, M., Finkelstein, S.~L., Ashby, M.~L.~N., et al.\ 2016, ApJ, 825, 5 \\
$\bullet$ Stern, D., Eisenhardt, P., Gorjian, V., et al. 2005, ApJ, 631, 163\\
$\bullet$ Tasnim Ananna, T., Treister, E., Urry, C.~M., et al.\ 2019, ApJ, 871, 240\\
$\bullet$ Tombesi, F., et al. 2019, {\it Astro2020 Science White Paper: ``Do Supermassive Black Hole Winds Impact Galaxy Evolution?''}\\
$\bullet$ The Lynx Team\ 2018, arXiv e-prints , arXiv:1809.09642\\
$\bullet$ Ueda, Y., Akiyama, M., Hasinger, G., Miyaji, T., \& Watson, M.~G.\ 2014, ApJ, 786, 104 \\
$\bullet$ Vito, F., Gilli, R., Vignali, C., et al.\ 2016, MNRAS, 463, 348 \\
$\bullet$ Vito, F., Brandt, W.~N., Yang, G., et al.\ 2018, MNRAS, 473, 2378 \\
$\bullet$ Volonteri, M., Reines, A.~E., Atek, H., Stark, D.~P., \& Trebitsch, M.\ 2017, ApJ, 849, 155 \\
$\bullet$ Weinberger, R., Springel, V., Pakmor, R., et al.\ 2018, MNRAS, 479, 4056 \\
$\bullet$ Yang, G., Chen, C.-T.~J., Vito, F., et al.\ 2017, ApJ, 842, 72 \\
$\bullet$ Yang, G., Brandt, W.~N., Darvish, B., et al.\ 2018, MNRAS, 480, 1022 \\
$\bullet$ Zappacosta, L., Piconcelli, E., Duras, F., et al.\ 2018, A\&A, 618, A28 \\

\end{document}